\documentstyle[aps,amssymb,preprint]{revtex}

\begin{document}
\author{B.K.CHAKRAVERTY}
\address{LEPES/CNRS, BP166, 25 Av. des Martyrs, F-38042 Grenoble Cedex 9}
\title{{\bf Quantum Phase Fluctuation in High T}$_{c}${\bf \ Superconductors}}
\date{July 1st, 1999}
\maketitle

\begin{abstract}
If phase coherence determines the superconducting transition temperature T$%
_{c}$ in the cuprate oxides, it is of great interest to understand the role
that dynamics of the phase fluctuation plays in bringing about depletion of
the superconducting condensate. We will show that a phase correlation
function can be calculated that allows us to describe depletion of
superconducting condensate as a result of the quantum fluctuation. In two
dimension, dynamic phase fluctuation or pair fluctuation gives rise to a
condensate depletion linear with temperature as T$\rightarrow O$ in
superconductors with nodes at the fermi surface.
\end{abstract}

Superconducting order parameter has an amplitude and a phase. In the ground
state the phase has to be blocked in all space. It has been suggested \cite
{chakraverty} that the superconducting transition temperature T$_{c}$ in the
oxide materials is related to the loss of long range phase coherence. Beyond
T$_{c}$ the system lives, in this model, unlike classical superconductors in
an ill understood mixture of Cooper pairs that have a nonzero amplitude and
a highly fluctuating phase. The dynamics of the phase fluctuation and
depletion of superfluid density of the superconducting state is the object
of this study.

We write the superconducting order parameter as $\psi \left( t\right) $
.where 
\begin{equation}
\psi \left( t\right) =\left| \psi \right| \exp \left( i\theta \left( x\text{,%
}t\right) \right)  \label{order parameter}
\end{equation}
where $\left| \psi \right| $ is the amplitude assumed constant in space and $%
\theta \left( x,t\right) $ is the phase assumed fluctuating in space and
time..The phase dynamics is contained in the simple hydrodynamic phase
fluctuation hamiltonian \cite{Ramakrishnan} 
\begin{equation}
H=\frac{1}{2}\kappa _{s}\left( \nabla \theta \right) ^{2}+\frac{1}{2\chi
_{\rho }}\left( \partial \rho _{s}\right) ^{2}+\frac{1}{2}\partial \rho
_{s}e^{*}\phi \left( t\right)  \label{hamiltonian}
\end{equation}
Here $\kappa _{s}$is the superfluid phase stiffness and $\chi _{\rho }$ is
the density susceptibility related to the density of states of paired
electrons, $\partial \rho _{s},$ a local density fluctuation that has long
range coulomb interaction with nonlocal potential $\phi \left( t\right) $
and $e^{*}$is the pair charge.We note that $\theta $ and the pairdensity $%
\rho _{s}$ are conjugate quantities so that all the dynamics of $\theta $ is
contained in the excess pair charge $e^{*}\partial \rho _{s}$ , the
subscript s meaning superconducting phase.We have the Josephson relation
governing the dynamics of the phase to induced local potential $\phi \left(
t\right) $ through 
\begin{eqnarray}
\frac{\partial \theta }{\partial t} &=&\frac{e^{*}}{\hbar }\phi \left(
t\right)  \nonumber \\
\nabla ^{2}\phi &=&e\left( 2\partial \rho _{s}-\rho _{qp}\right)
\end{eqnarray}
Here $\rho _{qp}$ is local quasiparticle population, if there is any . Let
us assume that the coulomb interaction given by the third term of the
hamiltonian (2) is zero. The resulting hamiltonian can be written as 
\begin{equation}
H_{o}=\int d\text{v }\left( \frac{\rho _{o}\nu _{s}^{2}}{2}+\frac{
c_{o}^{2}\left( \partial \rho _{s}\right) ^{2}}{2\rho _{o}}\right)
\end{equation}
where $\rho _{o}$is the condensate mass density, $\nu _{s}$is the superfluid
flow, proportional to $\triangledown \theta \left( x,t\right) $ and $c_{o}$
is the bare fermi velocity.The hamiltonian gives the elementary phase
excitation as the Boguylubov-Anderson mode with the well-known dispersion $%
w_{q}=cq$ where is $c$ =$\frac{c_{o}}{\sqrt{2}},$ in 2d . In order to obtain
the depletion of superfluid density due to dynamic phase fluctuation we use
the usual trick of adding to the phase hamiltonian (4) an amplitude
modulation term that represents an infinitesimal coupling to the medium , a
source term 
\begin{equation}
H^{\prime }=-\frac{1}{2}\eta \sum_{x}\left( \psi +\psi ^{*}\right)
\end{equation}
Where $\eta $ is the interaction between pairs and the medium which in the
end will be allowed to go to zero . The fluctuation of the longitudinal part
of the order paramter is defined through 
\begin{eqnarray}
\Delta ^{\prime }\left( x\right) &=&\Delta -\Delta _{o}  \nonumber \\
\Delta &=&\psi +\psi ^{*};\Delta _{o}=\psi _{o}+\psi _{o}^{*} \\
\Delta ^{\prime } &=&-\sqrt{\frac{n_{o}}{\Omega }}\sum_{q}\theta _{q}^{2} 
\nonumber
\end{eqnarray}
Here n $_{o}$ is the condensate density ($n_{o}\sim \frac{N_{h}}{2}$ where $%
N_{h}$ is the hole density in oxide materials), $\Omega $ is an atomic
volume in 3d (in 2dimension n$_{o}$ is the pair density per unit area while $%
\Omega $ is the area of an unit cell on the surface ), $\theta _{q}$is the
fourier transform of the phase fluctuation $\theta .$ The total fluctuation
hamiltonian reads 
\begin{equation}
H=H_{o}+H^{^{\prime }}
\end{equation}
This can be transformed in the fourier space as 
\begin{equation}
H\left( \eta \right) =\sum_{q}\left[ \frac{n_{o}q^{2}}{2m^{2}}\left| \theta
_{q}\right| ^{2}+\frac{mc_{o}^{2}}{2n_{o}}\left| \rho _{q}\right| ^{2}+\frac{
\eta }{2}\sqrt{\frac{n_{o}}{\Omega }}\left| \theta _{q}\right| ^{2}\right]
\end{equation}
The excitation spectra of the phase mode is significantly modified, due to
coupling $\eta $ and is given by $($after restoring $\hbar )$%
\begin{equation}
\varpi _{q}\left( \eta \right) =\hbar cq\left( 1+\frac{\eta m}{\hbar
^{2}n_{o}q^{2}}\sqrt{\frac{n_{o}}{\Omega }}\right) ^{\frac{1}{2}}
\end{equation}
At small momenta we now have a gap in the excitation spectra given by 
\begin{equation}
\varpi _{o}\left( \eta \right) =c\left[ \frac{\eta m}{n_{o}}\sqrt{\frac{n_{o}%
}{\Omega }}\right] ^{\frac{1}{2}}
\end{equation}
It is relatively straightforward to derive the relation for the condensate
density $n_{o}\left( T\right) $ at low temperature due to phase fluctuation.
By writing the grand canonical hamiltonian 
\begin{equation}
\hat{H}=H-\mu N
\end{equation}
we can show that \cite{negele} 
\begin{equation}
\left\langle \psi \right\rangle =\sqrt{n_{o}\left( T\right) \Omega }
=-\left\langle \frac{\partial \hat{H}}{\partial \eta }\right\rangle _{\eta
=o}=-\left| \frac{\partial G\left( \eta \right) }{\partial \eta }\right|
_{\eta =o}
\end{equation}
Here $G$ is the Gibb's free energy and is given by 
\begin{equation}
G\left( \eta ,T\right) =G\left( \eta ,o\right) +\frac{kT\Omega }{\left( 2\pi
\right) }\int qdq\text{ }\ln \left[ 1-\exp -\beta \varpi _{q}\left( \eta
\right) \right]
\end{equation}
This allows us to write in two dimension 
\begin{equation}
n_{o}\left( T\right) =n_{o}-\frac{\sqrt{n_{o}\Omega }}{\pi }
\int_{q}qdqN_{q}^{o}\left[ \frac{\partial \varpi _{q}\left( \eta \right) }{%
\partial \eta }\right] _{\eta =o}
\end{equation}
Here $N_{q}^{o}$ is the bose factor for the unperturbed energy given by 
\begin{equation}
N_{q}^{o}=\frac{1}{\exp \left( \frac{\hbar cq}{kT}\right) -1}
\end{equation}
Using equation (9) and substituting it into equation $\left( 14\right) $ ,
we obtain integrating equation $\left( 14\right) $ as $T\rightarrow 0,$ the
condensate density in two dimension 
\begin{equation}
\frac{n_{o}\left( T\right) -n_{o}}{n_{o}}=-\frac{1}{2\pi }\frac{mkT}{
n_{o}\hbar ^{2}}\simeq -\frac{1}{\pi }\frac{mkT}{N_{h}\hbar ^{2}}
\end{equation}
In this expression m is the electronic mass and $N_{h}$ is the total hole
density. Care has to be exercised in using the expression (16) as $n_{o},$
the pair density at T=0 may be considerably less than $N_{h}/2$ due to
quantum fluctuation which we calculate in the next section.In three
dimension the same calculation gives a $T^{2}$ diminution of the condensate
fraction as T$\rightarrow 0$.

The quantum phase fluctuation will be studied with an lattice hamiltonian
equivalent to that of expression (4) that we have used earlier $\left[
4\right] .$We have pointed out that almost all high T$_{c}$ cuprates are
characterised universally as domains of coherence length $\xi $ separated by
about the same distance from neighboring domains and that phase coherence
between the domains is assured by condensate pair transfer between them at a
rate given by the Josephson coupling energy J. This gives the discrete or
coarse-grained version of the hamiltonian of (4) as a Josephson lattice
hamiltonian 
\begin{equation}
H_{\theta }=-J\sum_{ij}\left( \cos (\theta _{i}-\theta _{j})\right) +\frac{%
\left( \partial \rho _{s}\right) ^{2}}{2\chi _{\rho }}
\label{josephsonhamiltonian}
\end{equation}
The energy increase for a single cooper pair charge $\partial \rho _{s}=\pm
2e$ is given by in this pure transfer process 
\begin{equation}
{\LARGE \epsilon }_{c}=\frac{\left( \partial \rho _{s}\right) ^{2}}{2\chi
_{\rho }}=\frac{2e^{2}}{\chi _{\rho }}  \label{Coulomb energy}
\end{equation}

$\epsilon _{c}$ is the quantum unit of energy increase for a single cooper
pair transfer.At $T=0$, the lattice hamiltonian is equivalent to a $d+1$
system and is given exactly by 
\begin{equation}
H_{\theta }^{d+1}=\frac{1}{g}\sum_{ij}\cos \left( \theta _{i}-\theta
_{j}\right)
\end{equation}
where a quantum coupling constant for quantum fluctuation can be defined
given by 
\begin{equation}
g=\sqrt{\frac{\epsilon _{c}}{J}}
\end{equation}
We have shown $\left[ 4\right] $ that beyond a critical coupling constant $%
g=g_{c},$ quantum fluctuation destroys the superconducting ground state and
drives the system insulating. We will derive a similar characteristic from
the phase correlation function below.

We define the phase correlation function at a point x by 
\begin{equation}
\Phi _{x}=\left\langle \exp \left[ i\theta \left( t\right) -i\theta \left(
0\right) \right] \right\rangle  \label{phasecorrelator}
\end{equation}

Since phase and charge are conjugate variables ,the charge operator $\rho $
induces transition from the ground state $\langle \Psi _{G}\mid $ to an
excited state $\langle \Psi _{E}\mid $ with a probability given by the
matrix element $\left| \left\langle \Psi _{E}\left| \rho \right| \Psi
_{G}\right\rangle \right| ^{2}$such that we have the sum rule 
\begin{eqnarray}
\sum_{E}\left( E-E_{G}\right) \left| \left\langle \Psi _{E}\left| \rho
\right| \Psi _{G}\right\rangle \right| ^{2} &=&{\LARGE \epsilon }_{c}
\label{sumrule} \\
&&  \nonumber
\end{eqnarray}

The phase correlator can be written as 
\begin{equation}
\left\langle \exp -i\frac{\left( E-E_{g}\right) t}{\hbar }\right\rangle
=\Phi _{x}  \label{Energy average}
\end{equation}

We will now calculate the phase correlator $\Phi $ using the lattice
hamiltonian (16). Considering $\theta $ as an operator we write the
commutator of $\theta \left( 0\right) $ and $\theta \left( t\right) $%
\begin{equation}
\left[ \theta \left( 0\right) ,\theta \left( t\right) \right] =\frac{i\hbar 
}{\chi _{\rho }}\sin \varpi _{o}t  \label{Commutator}
\end{equation}

where $\varpi _{o}$ is a characteristic frequency of the phase mode. We now
use the operator identity 
\[
\exp (A+B)=e^{A}e^{B}e^{-\frac{\left[ A,B\right] }{2}} 
\]
to obtain 
\begin{eqnarray}
&&\exp i\theta \left( 0\right) \exp -i\theta \left( t\right)  \nonumber \\
&=&\exp i\left[ \theta \left( 0\right) -\theta \left( t\right) \right] \exp
\left( \frac{i{\LARGE \epsilon }_{c}}{\hbar \varpi _{o}}\sin \varpi
_{o}t\right)
\end{eqnarray}

we note that at small time $\theta \left( t\right) \approx \theta \left(
o\right) $ and the correlator behaves like $\exp \left( i\frac{\epsilon _{c}t%
}{\hbar }\right) $ as it should.

This expression can be fed back into (\ref{Energy average}) to get after a
little algebra 
\begin{equation}
\left\langle \exp -i\frac{\left( E-E_{g}\right) t}{\hbar }\right\rangle
=\Phi _{x}=\sum_{n=-\infty }^{n=+\infty }P(n\hbar \varpi _{o})\exp -in\varpi
_{o}t
\end{equation}

The fourier transform P{\bf (}n$\hbar \varpi _{o})$ gives the probability
that $n$ quanta of phase oscillation mode $\varpi _{o}$ will be excited in
the medium.This is explicitly written in terms of the Bessel functions I$%
_{n} $%
\begin{equation}
P{\bf (}n\hbar \varpi _{o}{\bf )=}\exp -\left\langle \theta
^{2}\right\rangle {\bf I}_{n}\left( \frac{{\LARGE \epsilon }_{c}}{\hbar
\varpi _{o}\sinh \left( \frac{\hbar \varpi _{o}\beta }{2}\right) }\right)
\exp -\left( \frac{n\hbar \beta \varpi _{o}}{2}\right)
\label{Probability of quanta emission}
\end{equation}

Here $\exp -\left\langle \theta ^{2}\right\rangle $ is like the Debye-waller
factor and is given by the ratio of two characteristic enegies of the
problem 
\begin{equation}
\left\langle \theta ^{2}\right\rangle =\frac{{\LARGE \epsilon }_{c}}{\hbar
\varpi _{o}}=g  \label{quantum coupling constant}
\end{equation}

where a quantum coupling constant has been defined which tells us how many
quanta will be excited during a typical pair transfer so as to cause
dephasing and depletion of the superconducting condensate.The expression for 
{\bf P(0) gives us the probability of condensate fraction at a given T. }

This allows us to write for the spectral density at T=0 as 
\begin{equation}
A(\omega )=\left( \exp -g\right) \left( \sum_{n=0}^{n=\infty }\frac{g^{n}}{n!%
}\delta \left( \omega -\epsilon _{c}-n\varpi _{o}\right) \right)
\label{spectral weight}
\end{equation}
$.$The spectral density obey the following sum rules 
\begin{equation}
\int_{o}^{\infty }\omega A\left( \omega \right) d\omega =\epsilon _{c}
\end{equation}
\begin{equation}
\int_{o}^{\infty }\omega ^{m}A\left( \omega \right) d\omega =\left\langle
\omega ^{m}\right\rangle
\end{equation}

Expression(\ref{spectral weight}) is plotted in figure 1 for a variety of
values of g. The rapid depletion of spectral weight $A\left( \omega
=o\right) $ i.e that of superfluid density at around g=g$_{c}$ is to be
seen. The two quantum coupling constants of the expressions (28) and (20)
are the same if we identify the characteristic energy $\varpi _{o}$with the
average energy scale of the Josephson hamiltonian (19) as $\varpi _{o}=\sqrt{
\epsilon _{c}J};\varpi _{o}$ is just then a debye energy or a characteristic
pair fluctuation frequency. We notice from the figures how the spectral
weight goes from a poisson distribution to a gaussian as the coupling
constant increases.The gaussian spectral distribution indicates that the
ground state is coupled to a bath of continuum of excitations. We also note
that the spread of the spectral weight as g increases does not mean we are
in the excited state for we are still at T=0. If there were no interactions
(g=0) we will get a single delta function at $\omega $ = $\epsilon _{c}.$
For g$\neq 0$ we are still in the ground state but the coupled system has n
quanta of phase oscillation modes excited with a certain probability P$%
\left( n\hbar \varpi _{o}\right) .$The effective superfluid density at T=0
can be written as 
\begin{equation}
\frac{n_{o}}{N_{h}}\sim \exp -g
\end{equation}

We show in table {\bf 1 } this fraction for a variety of optimum doped
materials taken from \cite{tanner}.In the table the effective number of
holes per copper, n$_{eff}$ has been measured by summing the oscillator
strength N$_{eff}(\omega )$ upto a frequency of the order of charge transfer
band ($\sim 12000$ cm$^{-1}).$ As temperature is reduced, the increase in
area at low frequencies is compensated by a decrease in area at high
frequencies. {\it However, below T}$_{c},${\it there is a loss of area,
which must be coming mostly from the free carrier peak at }$\omega
\rightarrow 0${\it , }as the authors conclude{\it .}Tne number of
supeconducting carriers n$_{s}$ is obtained in two ways: from the London
penetration depth $\lambda _{L}$ and from the Drude peak, in the normal
state. The measured n$_{s}$ values are almost same as the number of Drude
carriers n$_{d}($these later are not shown here). Thus only a quarter to a
fifth of the total doping induced holes appear in the delta function !A
value of g$\approx 1$ can be taken indicating that even at optimum doping
considerable quantum fluctuation is going on. It is interesting to compare
this ratio of $\frac{n_{eff}}{n_{s}}$with Ornstein et al's \cite{tanner}
optical data where a bulk plasma frequency from band structure value was
estimated to be $\sim $ 3.5e.v compared to 1.4 e.v of the superfluid
condensate at $\omega $=0 for YBa Cuo and which reflects faithfully the
ratio observed in the table 1. 
\[
{\bf Table1} 
\]

\[
\begin{tabular}{|c|c|c|c|c|}
\hline
material & T$_{c}\left( K\right) $ & $\frac{n_{eff}}{cu}$ & $\frac{n_{s}}{cu}
$ & $\frac{n_{s}}{n_{eff}}\%$ \\ \hline
La$_{2}cuo_{4}$ & 40 & 0.15 & 0.028 & 19 \\ \hline
BiPbSr$cu_{2}o_{8}$ & 79 & 0.37 & O.O85 & 23 \\ \hline
BiSrCa$cu_{2}o_{8}^{c}\left( a\right) $ & 85 & 0.44 & 0.100 & 23 \\ \hline
BiSrCa$cu_{2}o_{8}^{c}\left( b\right) $ & 85 & 0.48 & 0.090 & 19 \\ \hline
YBa$_{2}cu_{3}o_{7}\left( a\right) $ & 91 & 0.44 & 0.096 & 22 \\ \hline
YBa$_{2}cu_{3}o_{7}\left( b\right) $ & 91 & 0.59 & O.125 & 21 \\ \hline
Tl$_{2}Ba_{2}cacu_{2}o_{8}$ & 110 & 0.54 & 0.115 & 21 \\ \hline
\end{tabular}
\]

The underlying physics of quantum phase fluctuation is quite similar to the
boson shake up process in the strong coupling polaron problem \cite
{ranninger}

We are now in a position to calculate the temperature coefficient of
depletion of superfluid density using expression(16) and (33) and
renormalising the temperature T by $\tau =\frac{T}{T_{c}}$ so that we get
the depletion coefficient in the dimensionless form as 
\begin{equation}
\frac{\partial \left( \frac{n-n_{o}}{n_{o}}\right) }{\partial \tau }=-\frac{%
\exp \left( g\right) mkT_{c}}{\pi N_{h}\hbar ^{2}}
\end{equation}
Using for high $T_{c}$ supercoductors the following numbers $%
T_{c}=100^{o}k,m=5$ electron mass, $N_{h}\approx 10^{14}/cm^{2}$ and $g=1,$
we obtain the coefficient as $\sim $ $-$ $O.56$ in reasonable agreement with
the values in the literature\cite{hardy}. The ratio on the right hand side
is $mkT_{c}/n_{s}$, which is really a constant as shown by Uemura's plot.  A
linear T decrease of superconducting condensate due to{\it \ static phase
fluctuation} has been suggested in the past \cite{stroud} and that has been
recently contested \cite{larkin} .There is also the theory of low energy
quasi-particle excitation \cite{lee} at the d-wave gap nodes explaining the
linear T-dependence but the numerical agreement seems to need considerable
fermi-liquid correction \cite{larkin} to fit the data.

\smallskip The overall treatment in this paper has assumed perfect screening
of charge fluctuation $\partial \rho _{s}$in the condensate so that plasma
mode is not excited. we can speculate that in d-wave superconductor this may
well be so due to possibility of low energy quasiparticle excitation as a
result of phase excitation. The latter shifts the quasiparticle energy due
to doppler effect \cite{degennes}of the local superfluid flow by an amount 
\[
\partial E_{qp}=\hbar k_{f}^{\rightarrow }v_{s}^{\rightarrow } 
\]
at the fermi surface i.e. at the nodal points. Here v$_{s}=-\left( \hslash
/2m\right) \triangledown \theta ,$ is the local superfluid velocity. The
quasiparticle population created due to this doppler shift is $\rho
_{qp}\sim \int_{o}^{\partial E_{qp}}f\left( E\right) N\left( E\right)
\partial E$ where $N\left( E\right) $, the quasiparticle density of states.
The integral picks up significant weight only because of d-wave nodes, where
the product $N(E)$ $f(E)$ is $\neq 0.$ The condition of perfect screening
(local charge neutrality) $\rho _{qp}\approx \partial \rho _{s}$ may be
easily satisfied so that long range coulomb interaction is suppressed. A
background quasiparticle density of the order of 10\% has been inferred from
microwave measurements on high $T_{c}$ oxides\cite{waldram}.The local charge
neutrality condition implies div J$_{s}+$div J$_{qp}=$ 0 , where J's are
superconducting and quasiparticle currents respectively. For intense local
phase gradient , the more stringent well-known criterion of total ( although
local), destruction of condensate,due to quasiparticle backflow current
applies with J$_{s}^{crit}=$ -J$_{qp}^{crit}.$

Figure 1: spectral density $A\left( \omega \right) $ as a function of phase
quanta $n$. a) coupling constant $g=1$; b) $g=3$; c) $g=8$.


\begin{references}
\bibitem{chakraverty}  B.K. Chakraverty, A. Taraphder, M. Avignon, Physica
C235-240, 2323 (1994); V.J. Emery and S. Kivelson, Nature 374 , 434 (1995)

\bibitem{Ramakrishnan}  T.V. Ramakrishnan, Physica Scripta T27, 24 (1989);
S. De Palo, C. Castellani, C. Di Castro and B.K. Chakraverty,
cond-mat/9901305, accepted in Phys Rev B

\bibitem{negele}  John W.Negele and Henri Orland ``Quantum many particle
system'' Addison Wesley (1988) p.108

\bibitem{doniach}  B.K. Chakraverty and T.V. Ramakrishnan, Physica C 282,
290 (1997); S. Doniach, Phys Rev B24, 5063 (1981)

\bibitem{tanner}  D.B. Tanner et al, Physica B 244, 1 (1998); Joseph
Orenstein et al, Phys Rev B42, 6342 (1990)

\bibitem{ranninger}  J. Ranninger and A. Romano, Phys.Rev.Lett, 80(1998); S.
Ciuchi, F. de Pasquale, S. Fratini and D. Feinberg, Phys.Rev.B 56, 4494
(1997)

\bibitem{hardy}  W.N. Hardy et al, Phys.Rev.Lett. 70, 3999 (1993)

\bibitem{stroud}  E.Roddick and D.Stroud, Phys. Rev. Lettr., 74, 1430 (1995)

\bibitem{larkin}  A.J.Mills, S.M.Girvin, L.B.Ioffe and A.I.Larkin,
Cond-mat:9709222

\bibitem{lee}  P.A.Lee and X.G.Wen, Phys.Rev.Lett. 78 , 4111 (1997)

\bibitem{degennes}  P.G. de Gennes, ``Superconductivity of Metals and
Alloys'' (Addison-wesley,N.Y.1992), p.144; M. Franz and A.J. Millis,
Phys.Rev.B 58, 14572 (1998)

\bibitem{waldram}  J.R. Waldram et al, Phys Rev B 55, 3222 (1997)Phys Rev B
55, 3222 (1997)(1997)
\end{references}
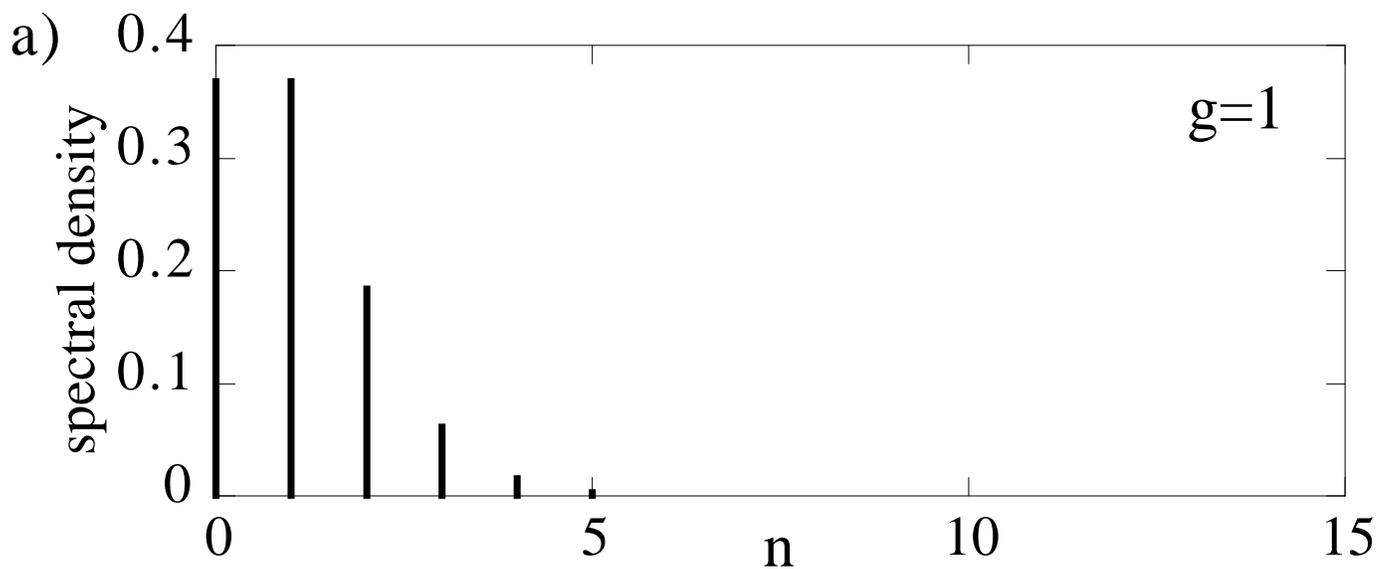
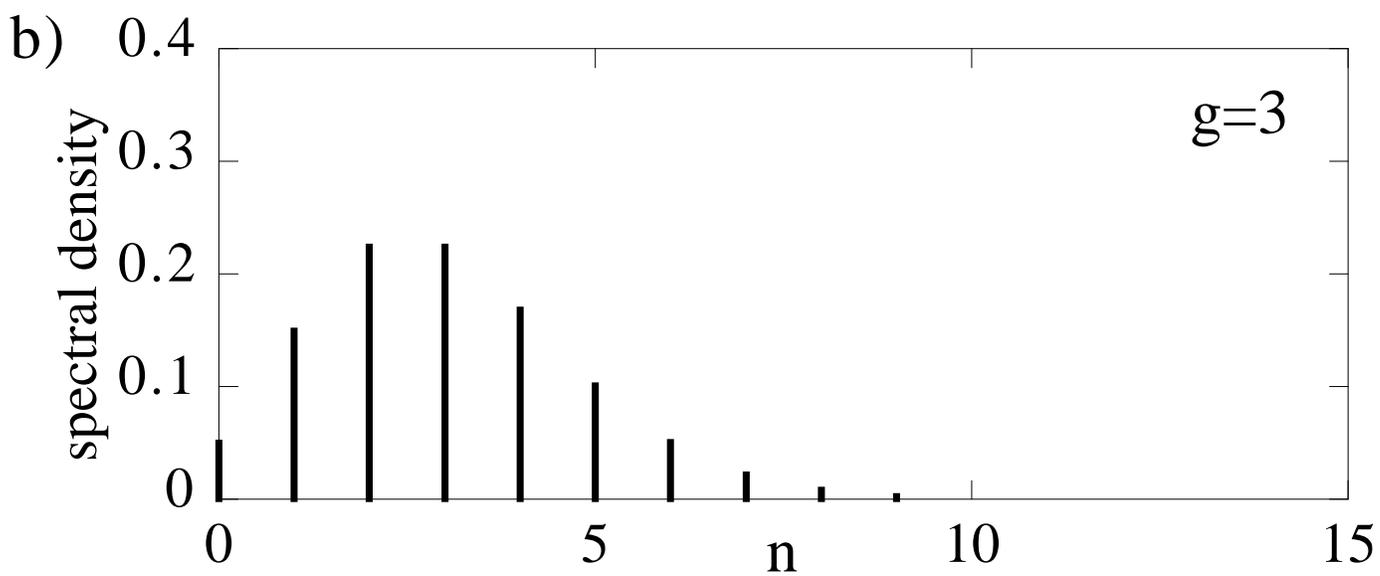
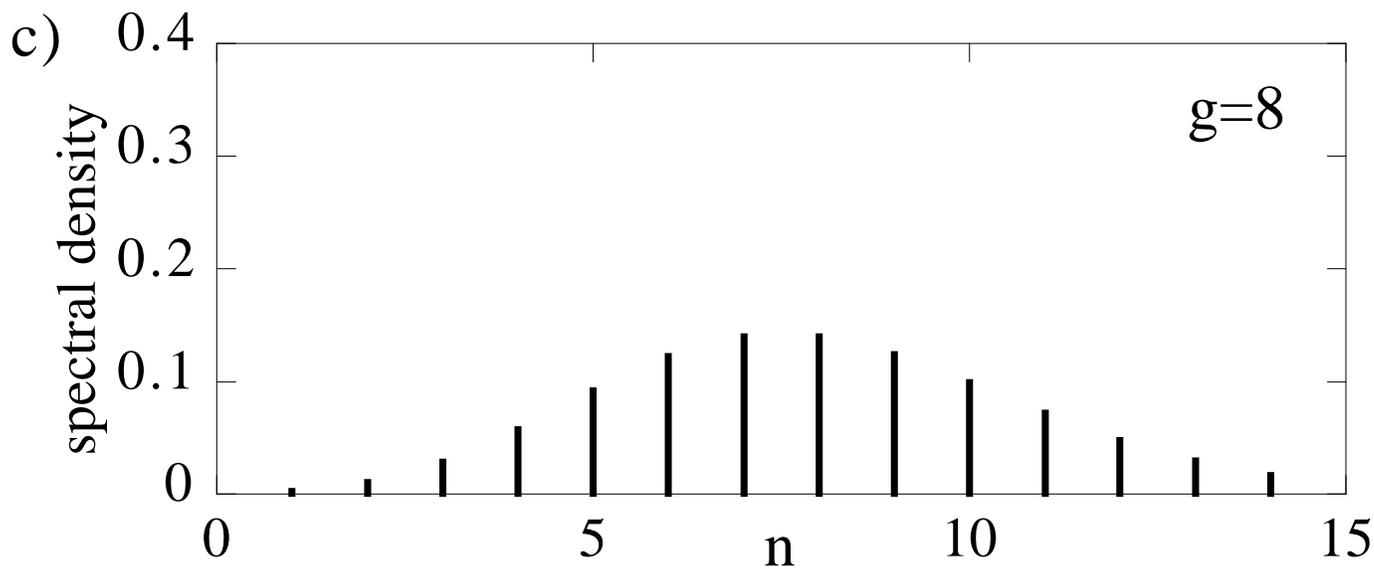
\end{document}